\documentclass[conference, 10pt]{IEEEtran}

\usepackage{algorithmic}
\usepackage{framed,multirow}
\usepackage{latexsym}
\usepackage[utf8]{inputenc}
\usepackage{url}
\usepackage{xcolor}
\usepackage{colortbl}
\definecolor{newcolor}{rgb}{.8,.349,.1}
\usepackage[normalem]{ulem}
\usepackage{cite}
\usepackage{makecell}
\usepackage{amsmath,amsfonts,amssymb}
\usepackage{svg}
\usepackage{multirow}
\usepackage{tabularx}
\usepackage{graphicx}
\usepackage{textcomp}
\usepackage{listings}
\usepackage{caption}
\usepackage{pdfpages}
\usepackage{hyperref}
\usepackage{units}
\usepackage{float}

\def\BibTeX{{\rm B\kern-.05em{\sc i\kern-.025em b}\kern-.08em
    T\kern-.1667em\lower.7ex\hbox{E}\kern-.125emX}}

\definecolor{accred}{rgb}{0.8,0.1,0.2}
\definecolor{idorange}{rgb}{0.9,0.5,0.0}
\definecolor{ivectblue}{rgb}{0.0,0.1,0.6}
\definecolor{highlightgreen}{rgb}{0.0,0.8,0.0}

\lstdefinestyle{f90}{
  language=[90]Fortran,
  basicstyle=\small,
  commentstyle=\color{green},
  morecomment=[l][\color{gray}\textit]{! },
  morecomment=[l][\color{accred}\bfseries]{!\$acc },
  morecomment=[l][\color{accred}\bfseries]{!\$omp },
  emph={igaus,inode,idime,jdime,id1,id2},
  emphstyle={\color{idorange}},
  emph={[2]VECTOR_DIM,ivect},
  emphstyle={[2]\color{ivectblue}},
  moredelim=[is][\color{accred}\bfseries\sout]{[-}{-]},
  moredelim=[is][\color{highlightgreen}\bfseries]{[!}{!]},
  moredelim=[is][\color{gray}\textit]{[/}{/]}
}

\lstdefinestyle{asm}{
  language=[x86masm]Assembler,
  basicstyle=\small,
  keywordstyle=\bfseries,
  keywords={st, ld, add},
  commentstyle=\textit,
  morecomment=[l][\textit]{do},
  morecomment=[l][\textit]{end},
  emph={r1, r2},
  emphstyle={\color{idorange}}
}

\newfloat{lstfloat}{h}{lop}
\floatname{lstfloat}{Listing}

\begin{document}

\title{Alya towards Exascale: Optimal OpenACC Performance of the Navier-Stokes Finite Element Assembly on GPUs}

\author{\IEEEauthorblockN{Herbert Owen\IEEEauthorrefmark{1}, Dominik Ernst\IEEEauthorrefmark{2} \\ (\textit{Both authors contributed equally to this work}) \\ Thomas Gruber\IEEEauthorrefmark{2}, Oriol Lemkuhl\IEEEauthorrefmark{1}, Guillaume Houzeaux\IEEEauthorrefmark{1}, Lucas Gasparino\IEEEauthorrefmark{1} and Gerhard Wellein\IEEEauthorrefmark{2}} \\

\IEEEauthorblockA{\IEEEauthorrefmark{1}Barcelona Supercomputing Center (BSC)\\
Barcelona, Spain \\
Email: herbert.owen@bsc.es} \\
\IEEEauthorblockA{\IEEEauthorrefmark{2}National High Performance Computing Center NHR@FAU \\ Friedrich-Alexander-Universität Erlangen-Nürnberg, Germany \\
Email: dominik.ernst@fau.de}
}

\maketitle

\begin{abstract}
This paper addresses the challenge of providing portable and highly efficient code structures for CPU and GPU architectures. We choose the assembly of the right-hand term in the incompressible flow module of the High-Performance Computational Mechanics code Alya, which is one of the two CFD codes in the Unified European Benchmark Suite. Starting from an efficient CPU-code and a related OpenACC-port for GPUs we successively investigate performance potentials arising from code specialization, algorithmic restructuring and low-level optimizations.

We demonstrate that only the combination of these different dimensions of runtime optimization unveils the full performance potential on the GPU and CPU. Roofline-based performance modelling is applied in this process and we demonstrate the need to investigate new optimization strategies if a classical roofline limit such as memory bandwidth utilization is achieved, rather than stopping the process.
The final unified OpenACC-based implementation boosts performance by more than 50x on an NVIDIA A100 GPU (achieving approximately 2.5 TF/s FP64) and a further factor of 5x for an Intel Icelake based CPU-node (achieving approximately 1.0 TF/s FP64).

The insights gained in our manual approach lays ground implementing unified but still highly efficient code structures for related kernels in Alya and other applications.  These can be realized by manual coding or automatic code generation frameworks.
\end{abstract}


\section{Introduction}

Annotation based GPU programming approaches like OpenACC have been instrumental in the porting of large code bases with long development histories.
While the approach allows to quickly get to a GPU enabled unified code base, the efficiency of the ported code can be sub optimal.

Such a GPU port has been done for Alya, a Multiphysics HPC code part of the UEABS (Unified European Applications Benchmark Suite)~\cite{UEABS}.
This initial GPU port relies on a vectorized, unified implementation for both GPU and CPU~\cite{BORRELL}.
Using this initial port, a single A100 GPU is about $4-5\times$ factors slower than an Icelake based CPU node.

To improve on the unsatisfactory performance, this work initially pursues a more flexible, non-unified GPU implementation.
The presented measures to tackle the identified problems make the GPU version more than $50\times$ faster than it was before.
We believe that the observed anti patterns and solutions can be transferable to other code bases with a similar development history.

We show that the different targets, GPU and CPU, have very different execution characteristics and problems, and optimization needs to be target specific.
At the same time, we find that most measures that make up the new, GPU friendly code base are compatible and even beneficial for the CPU as well, for an improvement of the CPU execution of more than $5\times$.
In the end, while analysis and optimization needs to be individually done for each target, this does not preclude the feasibility of a unified code base that is a reasonable performance compromise for both targets.
This also shows that implementations that would already be considered well optimized, because they score well by commonly used metrics like IPC (\emph{Instructions Per Cycle}) and core scaling for the CPU path, or because they reach a high portion of the roofline limit, like the GPU path, can still have considerable optimization headroom.

The application and the algorithm are described briefly in Section~\ref{FEAlgorithm}, and Section~\ref{machines} introduces the test case and hardware used.
In Section~\ref{statusquo}, we present the initial implementation, analyze the performance and find the reasons for the low GPU performance.
Section~\ref{improvements} then introduces three different measures that help to overcome the identified problems, and benchmarks and analyzes their impact.

\begin{figure}
\centering
    \includegraphics[width=8cm]{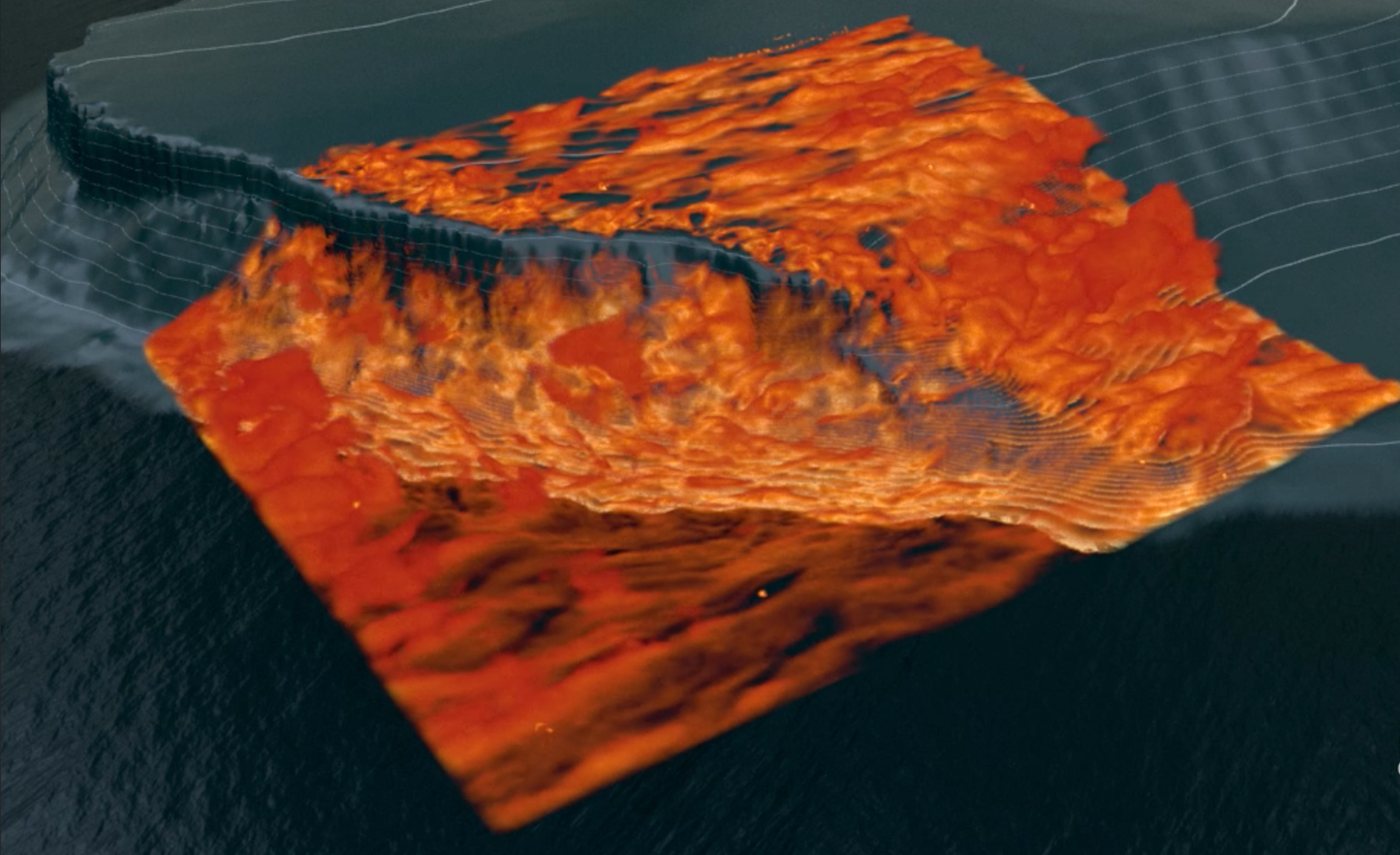}
    \caption{Volume rendering of the flow over the Bolund cliff, the test benchmark data set used in this paper}
    \label{fig:Bolund}
  \end{figure}

\section{Finite Element Assembly in CFD} \label{FEAlgorithm}

Understanding turbulent flow dynamics is crucial for improving energy efficiency since approximately 10\% of the world's energy use is spent overcoming turbulent friction.

A literature review on efficient finite element discretization of Navier Stokes equations on GPUs indicates that most recent approaches rely on high-order schemes. Kolev et al. \cite{kolev} state that the use of matrix-free high-order finite element methods is one of the few viable approaches to achieve high performance on GPUs. Moin and Verzicco \cite{MOIN2016242} argue that the benefits of high-order methods are significantly reduced in complex turbulence simulations and advocate using energy-preserving second-order methods, such as the one used by Alya. Considering the substantial evolution of GPUs in the last ten years (for example, in the amount of available memory), available literature on GPU optimization of low-order finite element approaches for flow problems \cite{Cecka_Lew} is outdated and of reduced significance for current GPUs.

For incompressible \emph{Large Eddy Simulations} (LES) using a fractional step scheme with explicit time discretization for momentum, the main computational kernels are the assembly of the \emph{Right Hand Side} (RHS) and the solution of a linear system of equations for the pressure. In our experience with Alya, the assembly of the RHS is the most time-consuming step, taking up to $80\%$ of the total time. Since we use small time steps to resolve all the energy-containing scales, the solution of the linear system for the pressure is usually not computationally demanding. The reader can find a detailed explanation of the approach in \cite{LEHMKUHL201951}.
In this work, we shall concentrate only on the RHS assembly routine, and we plan to use external linear algebra libraries as to solve the linear system of equations for the pressure on the GPU, such as AMG4PSBLAS \cite{doi:10.1137/20M134914X}.

While Alya can use mixed meshes, including tetrahedral, hexahedral, prismatic, and pyramidal elements, this paper focuses only on tetrahedral elements.
This is not a considerable restriction since mixed meshes can easily be partitioned to contain only tetrahedral elements with commercially available meshing tools. Some finite element codes \cite{doi:10.2514/6.2013-373} only work with tetrahedra.


\section{Measurements and Machines}\label{machines}

\subsection{Case Description}

We use a LES of the flow for the Bolund benchmark, one of the best-known benchmarks for Atmospheric Boundary Layer flow over complex terrain.
A volume rendering of the instantaneous velocity field is shown in Figure~\ref{fig:Bolund}.
A tetrahedral finite element mesh with 5.6 M nodes and 32 M elements has been used.
Measured values are the median out of multiple runs.
Since measurement variances are usually low, error bars have been omitted in most cases.

\subsection{CPU measurements details}

The CPU measurements are done on a dual socket \emph{Intel Xeon Platinum 8360Y ``Icelake''} system with $2\times36$ cores from \emph{NHR@FAU}'s CPU cluster \emph{Fritz}~\cite{fritz}.
We use the tool \emph{likwid}~\cite{likwid2} to measure the hardware metrics in Table~\ref{tab:metricscpu} using hardware performance counters.

Alya has been compiled using \lstinline[style=f90]{VECTOR_DIM=16} and the compiler options \verb!-xCORE-AVX512!, \verb!-mtune=icelake-server! and \verb!-qopt-zmm-usage=high! to make Intel's classic Fortran compiler \emph{ifort} in version 2021.4 use AVX512 instructions with their full width, which results in single core speedups in the range of $10-30\%$.
Turbo mode is active for all measurements.

Additionally, we use \verb!likwid-bench! to measure a single socket load bandwidth \footnote{\lstinline{likwid-bench -t load -W S0:1GB:36}} of \unitfrac [179]{GB}{s}, as well as a single socket peak floating point rate \footnote{\lstinline{likwid-bench -t peakflops_avx512_fma -W S0:4MB:36}} of \unitfrac[2705]{GFlop}{s} using AVX512, which results in a machine arithmetic intensity of \unitfrac[15]{Flop}{B}.

Alya uses pure MPI for core level parallelism and relies on auto vectorization to utilize SIMD capabilities.

\subsection{GPU measurement details}

All GPU measurements have been made using a 40GB SXM4 A100 GPU from the \emph{NHR@FAU}'s GPU cluster \emph{Alex}~\cite{alex}.
We use NVIDIA's Nsight Compute to measure the hardware metrics in Table~\ref{tab:metrics} using hardware performance counters.

In \cite{BORRELL}, it was found that for GPUs the optimal \lstinline[style=f90]{VECTOR_DIM} was 196,608.
We use a larger value of 2048k, which does not change the performance of the baseline, but is significantly faster for the more optimized versions.

The \emph{nvfortran} compiler from NVIDIA's HPC toolkit in version 21.11 is used with \verb!-acc!, \verb!-ta=tesla:cc80!, \verb!-fast!, \verb!-Mnoopenmp!, \verb!-Munroll!, and \verb!-Mnoidiom! as the performance relevant compiler options.

A measured Scale kernel memory bandwidth of \unitfrac[1381]{GB}{s} from \cite{cudabenches} and a double precision floating point execution rate of \unitfrac[9.7]{TFlop}{s} results in a machine arithmetic intensity of the GPU of \unitfrac[7]{Flop}{B}.

\section{Status Quo: Vectorized Data Structures}\label{statusquo}

\begin{lstfloat}[t]
\begin{lstlisting} [style=f90]
real :: temp(VECTOR_DIM)
real :: output(VECTOR_DIM, 3)

output(:,:) = 0

!other contributions to output
[...]

!compute and store something in temp
temp(:) = [...]

do i = 1,3
  output(:, i) = output(:, i) + temp(:)
end do

[...]

\end{lstlisting}
  \caption{Example of how the CPU path would implement the addition of a scalar contribution stored in temp to a 3 element output array for \lstinline[style=f90]{VECTOR_DIM} elements. Compare with GPU path in Listing~\ref{lst:gpupath}} \label{lst:cpupath}
\end{lstfloat}

The RHS assembly function is written in a vectorized style, with each function call performing the computation for a vector of \lstinline[style=f90]{VECTOR_DIM} elements simultaneously.
While standard finite element codes perform a loop over all elements one by one, the vectorized approach loops over groups of elements \cite{BORRELL}.

Listings~\ref{lst:cpupath} and \ref{lst:gpupath} show a made up example for this style, which computes an output array with 3 elements from a scalar intermediate value \lstinline [style=f90]{temp}.
All results and intermediates, scalars or arrays, are allocated with an extra dimension of length \lstinline[style=f90]{VECTOR_DIM} to hold the each element's data.
That extra dimension is the first, innermost dimension, so that the data of different elements is stored interleaved.
That way, all loads and stores to intermediate and result values can be done with vectorized, coalesced load and store instructions, suitable for both CPU SIMD and GPU warp execution.

However, we observe that they also inhibit the ability of both tested compiler's (NVIDIA nvfortran for GPUs and Intel's ifort on CPUs) ability to analyze and transform the code.
Often, a temporary value can be kept in a register for the duration of its short lifetime.
Since the intermediate results will not be used anymore after its last use, storing it to memory is unnecessary.
Instead, both tested compiler do allocate memory for the temporary values and store the intermediate values to memory.
Even worse, we observe that both compilers implement statements like \lstinline[style=f90]{temp(:) = temp(:) + ...} as separate load/compute/store sequences.
Even for zero initialization, the compilers emit the store of a zero to memory, just to reload the zero from memory a few instructions later.
Arithmetic transformations that eliminate intermediate expressions are inhibited, because the compiler has to assume they are a desired byproduct that has to be stored to memory.

The RHS assembly routine in its original form uses 32 arrays with a total of 430 double precision values per element.
The high amount of intermediates far exceeds the amount of data that has to be read as inputs as well as the computed results.

For both compilers, the generated assembly contains a huge amount of load and store instructions and missed arithmetic transformations that could have enabled common subexpression elimination.
However, the consequences on performance are different depending on the target platform, CPU or GPU.

\paragraph{CPU Path} In the default version currently available in Alya, the Fortran preprocessor is used to create a CPU or GPU specific version from a unified codebase.
The version for the CPU uses a macro to insert Fortran array expressions for each computation, as in Listing~\ref{lst:cpupath}, showing an example typical for the CPU path.

That way, each statement in the code computes on a vector of \lstinline[style=f90]{VECTOR_DIM} elements.
Choosing a small multiple of the SIMD length for the vector lengths enables the auto vectorizer to use SIMD instructions as well as some loop unrolling.
We found \lstinline[style=f90]{VECTOR_DIM=16} to be fastest for both AVX256 and AVX512.
For this size, the data volume of the arrays allocated in the function for intermediate values is less than $100\;kB$, so that most arrays are covered by the L1 and L2 cache.
For larger vector lengths, the increased unrolling would not beneficial any more, and the combined data volume of all accessed data and intermediate results for all vector lanes would reduce the effectiveness of the L1 cache.

\definecolor{myc0}{rgb}{0.85,0.92,0.97}
\definecolor{myc1}{rgb}{0.96,0.96,1.0}
\definecolor{myc2}{rgb}{0.90,0.90,0.95}
\definecolor{myc3}{rgb}{1.0,0.92,0.82}
\definecolor{myc4}{rgb}{0.9,0.99,0.9}
\definecolor{myc5}{rgb}{0.99,0.99,0.98}
\definecolor{myw}{rgb}{1.0,1.0,1.0}
\newcolumntype{a}{>{\columncolor{myc0}}r}
\newcolumntype{d}{>{\columncolor{myc1}}r}
\newcolumntype{g}{>{\columncolor{myc2}}r}
\newcolumntype{k}{>{\columncolor{myc3}}r}
\newcolumntype{n}{>{\columncolor{myc4}}r}

\def\arraystretch{1.5}

\begin{table} [t]
  \centering
  \begin{tabular}[width=\textwidth]{  r | r |  r | r   }

                   \multirow{2}{4em}[-1.4em]{  \large{\textbf{CPU}}}  & \textbf{B}  &  \textbf {RS} & \textbf{RSP}    \\ \cline{2-4}
                     & \makecell[l]{\small{\textbf{B}aseline}}
                     & \makecell[l]{\scriptsize{+ \textbf{S}pecialization} \\ \scriptsize{+ \textbf{R}estructuring}}
                     & \makecell[l]{\scriptsize{+ \textbf{R}estructuring} \\ \scriptsize{+ \textbf{S}pecialization} \\ \scriptsize{+ \textbf{P}rivatization}} \\ \hline
  \multicolumn {4}{l} {operations per element} \\
  \rowcolor{myc0} load/store            &    6055         &   2516              &  639     \\ 
  \rowcolor{myc5} FP                    &    6316          &   1760              &  1249     \\ \hline
  \multicolumn {4}{l} {data volumes, Bytes per element } \\
  \rowcolor{myc0} L1 volume     &  48'440  &    20'128  & 5'112   \\
 \rowcolor{myc0} L1 effectiveness    &  $74\%$ &    $94\%$ & $82\%$ \\
  \rowcolor{myc5} L2/L3  volume &  12'716  &     1120  &    932   \\
 \rowcolor{myc5} L2/L3  effectiveness  &  $98\%$ &     $80\%$ &   $74\%$  \\
  \rowcolor{myc0} DRAM                  &    261           &    218              &    241           \\ \hline
  \rowcolor{myc5} GFlop/s, 1c           &   13.8           &    11.9             &   14.2          \\
  \rowcolor{myc5} GB/s, 1c              &   0.53           &     1.3             &    2.5         \\
 \rowcolor{myc0} runtime 1c, ms        &   44047          &   15429             &   8400        \\
  \rowcolor{myc0} runtime 71c, ms       &   785            &    244              &    122
\end{tabular}
\vspace{1mm}
\caption{CPU performance counter and performance measurements. 1 FMA = 2 Flop.  \emph{Operations per element} is (measured) \emph{executed instructions} $\times$ \emph{SIMD length} / \emph{element count}. \emph{L2/L3/DRAM volume} is measured with performance counters using LIKWID. \emph{L1 Volume} is $load/store \;operations / 2 \times 8$ (the factor ).  \emph{Cache effectiveness} is the percentage of cache traffic that hits in the cache. }

\label{tab:metricscpu}
\end{table}

The \textbf{B} or baseline column in Table~\ref{tab:metricscpu} has measurement data that shows high cache effectiveness:
$74\%$ of the volume requested from the L1 cache is served by the L1 cache and only $100\%-74\% = 26\%$ has to be fetched from the L2 cache.
Especially the unnecessarily repeated loads and stores of the same values have high L1 cache hit rates.
After accounting for the fact that the use of 256bit split loads (the compiler emits two 256bit loads to load a single 512bit wide register) doubles the number of vector load/store instructions, a similar number of load/store operations as floating point operations are executed for each element.
This is the result of not keeping operands in registers but repeatedly reloading them from memory.

Combined with the even higher combined L2/L3 cache effectiveness of $98\%$, few of the high number of load and store instructions require data from DRAM.
For each element, \unit[6316]{Flop} are executed and \unit[261]{B} miss all cache levels and are transferred between DRAM and CPU, resulting in a code arithmetic intensity of \unitfrac[24]{Flop}{B}, which is above the machine arithmetic intensity of \unitfrac[15]{Flop}{B} measured in Section~\ref{machines}.

Even though the code would thus be classified as compute bound by the roofline model~\cite{roofline} on the CPU, the floating point operation execution rate of \unit[13.8]{GFlop/s} is only about $20\%$ of the previously measured AVX512 peak.

\paragraph{OpenACC/GPU path}

\begin{table*}
\begin{tabular}[width=\textwidth]{ r r | r | r | r | r| r  }

  & \multirow{2}{4em}[-1.4em]{\large{\textbf{GPU}}}   & \textbf{B}  & \textbf{P} & \textbf{RS} & \textbf{RSP} & \textbf{RSPR}   \\ \cline{3-7}
    &                 & \makecell[l]{\footnotesize{  \textbf{B}aseline} \hspace{5mm}  }
                     & \makecell[l]{\footnotesize{+ \textbf{P}rivatization}\hspace{3mm} }
                     & \makecell[l]{\footnotesize{+ \textbf{S}pecialization} \\ \footnotesize{+ \textbf{R}estructuring}}
                     & \makecell[l]{\footnotesize{+ \textbf{R}estructuring} \\ \footnotesize{+ \textbf{S}pecialization} \\ \footnotesize{+ \textbf{P}rivatization}  }
                     & \makecell[l]{ \\[-1ex] \footnotesize{+ \textbf{R}estructuring} \\ \footnotesize{+ \textbf{S}pecialization} \\ \footnotesize{+ \textbf{P}rivatization} \\ \footnotesize{+ \textbf{R}estructuring} \\[1ex]}  \\ \hline

\rowcolor{myc0} \cellcolor{myw}  &   global load/store           &  6218  &  483  &  960  &    50  &  71   \\
 \rowcolor{myc5} \cellcolor{myw}   & local load/store           &    24  & 2593  &    0   &   71  & 30   \\
    \rowcolor{myc0}  \multirow {-3}{2cm} {\cellcolor{myw}\normalsize{operations  per element}} &
                    floating point              &  6293  & 6148  &  1663   &  1391  & 1333  \\ \hline
 \rowcolor{myc5}\cellcolor{myw} & L1  (effectiveness)           & 49'936 ($29\%$) & 24'616  ($3\%$) &  7680 ($60\%$) &  968  ($0\%$)   &  808 ($0\%$)  \\
 \rowcolor{myc0}\cellcolor{myw} & L2 (effectiveness)            & 35'507 ($34\%$) & 23'837 ($21\%$) &  3052 ($61\%$) &  1304  ($66\%$) &  968 ($84\%$) \\
  \rowcolor{myc5} \multirow {-3}{2.2cm} {\cellcolor{myw}\normalsize{data volume \\ Byte / element }}
 & DRAM          & 23'331          & 18'721          &  1170          &  442            &   150         \\ \hline
 \rowcolor{myc0}\cellcolor{myw} & registers                   &   255  &  255  &   184  &  148  &  128  \\
\rowcolor{myc5}  \cellcolor{myw} & GFlop/s                     &    163 &   393 &   829  &  2020 &  2575 \\
 \rowcolor{myc0} \cellcolor{myw} & GB/s                        &    608 &  1200 &    583 &  646  &  289  \\
\rowcolor{myc5} \cellcolor{myw}  & runtime in kernels, ms      &   3773 &  1536 &  197   &  68  &  51  \\
\end{tabular}
\vspace{2mm}
\caption{GPU performance counter and performance measurements. 1 FMA = 2 Flop.  \emph{Operations per element} is (measured) \emph{executed instructions} $\times$ \emph{warp length} / \emph{element count}. \emph{L2/DRAM volume} is measured with performance counters using \emph{Nsight Compute}. The \emph{L1 Volume} is $load/store \;operations \times 8$. \emph{Cache effectiveness} is the percentage of cache traffic that hits in the cache. }
\label{tab:metrics}
\end{table*}

\begin{lstfloat}
\begin{lstlisting} [style=f90]
real :: temp(VECTOR_DIM)
real :: output(VECTOR_DIM, 3)

!$acc parallel loop create(temp)
do ivect = 1,VECTOR_DIM
  output(ivect,:) = 0

  !other contributions to output
  [...]

  !compute and store something in temp
  temp(ivect) = [...]

  do i = 1,3
    output(ivect, i) = &
        output(ivect, i) + temp(:)
  end do
  [...]
end do
\end{lstlisting}
  \caption{Example of how the GPU path would implement the addition of a scalar contribution stored in temp to a 3 element output array for \lstinline[style=f90]{VECTOR_DIM} elements. Compare with CPU path in Listing~\ref{lst:cpupath} } \label{lst:gpupath}
\end{lstfloat}

To match the OpenACC programming model, the GPU version adds a loop over all vector elements around the whole function body.
The Fortran array expressions are replaced by the loop index \lstinline[style=f90]{ivect} instead.
An OpenACC \lstinline[style=f90]{!$acc parallel loop} statement maps the individual loop iterations to the GPU threads, as illustrated in Listing~\ref{lst:gpupath}.
Because contiguous sets of GPU threads execute and load data together, the interleaved storage of all temporary arrays also leads to uniform stride, coalesced memory accesses on the GPU.

One wave of concurrently executing threads on a modern GPUs contains in the range of $10^{6}$ threads, which is why we set the vector length \lstinline[style=f90]{VECTOR_DIM}, which determines the thread count for one GPU kernel launch, to 2048k in this paper in order to launch multiple waves with each kernel launch.
The data volume of all the intermediates of all concurrently running threads consequently comprises a much larger data volume than on the CPU.
Even at low occupancy, the data volumes exceed both the capacity of the per SM 192kB L1 cache, and the device wide 40MB L2 cache of the A100.

The measurements in the \textbf{B} or baseline column in Table~\ref{tab:metrics} show the consequences:
Both the L1 and the L2 cache have an effectiveness of only around $30\%$.
Each element requires 6293 floating point operations and the transfer of \unit[23331]{B} from and to DRAM, resulting in a code arithmetic intensity of about \unitfrac[$\frac{1}{3}$]{Flop}{B}.
Because this value is far below the machine intensity of \unitfrac[7]{Flop}{B} computed in Section~\ref{machines}, the roofline model would classify the execution of the RHS assembly on the GPU as memory bound, which contrasts with the compute boundedness of the CPU execution.

Despite being memory bound, the baseline version of the GPU cannot fully utilize the memory bandwidth at only about \unit[608]{GB/s} out of about \unit[1400]{GB/s}.
The short load/compute/store cycles offer little \emph{Memory Instruction Level Parallelism} (ILP) to batch together multiple cache misses to offset a GPU's high cache latencies.
This is exacerbated by low occupancy, caused by the compiler allocating the maximum 255 registers.

At \unit[3773]{ms} kernel execution time versus \unit[785]{ms}, the baseline version of the RHS assembly runs about $5\times$ slower on a single A100 GPU than on 71 CPU cores.

\section{Improvements}\label{improvements}

In conclusion, the large amount of intermediate values combined with the semantics of globally allocated temporary arrays lead to inflated memory volumes and low ressource utilization ($2\%$ of the FP peak) on the GPU.

We have categorized the measures to solve these problems in three categories:

\begin{itemize}
  \item \textbf{R}estructure which values are computed at what time and in which order.
  \item \textbf{S}pecialize, i.e., giving up some generality
  \item \textbf{P}rivatize the intermediate result arrays instead of allocating large global vectors.
\end{itemize}

To name different variants of the source code, we use combinations of the letters \textbf{R}, \textbf{S}, \textbf{P} to designate combinations of the different measures, as well as \textbf{B} for the baseline variant.

One optimization, that was not pursued, is the vectorization across the unknowns in each element, which is a common implementation pattern for finite element codes.
This could work in reducing the number of intermediates, as each vector lane would do fewer computations.
It would however, also lead to redundant computations or lane crossing communication, which would be more difficult to implement well for both CPU and GPU.

\subsection{\textbf{R}estructuring: Intermediate Value Lifetime Optimizations}\label{lifetimeopts}

During the RHS momentum assembly for each element, the RHS corresponding to all nodes in the element consists of contributions from different physical terms.
The most concise and human-readable way of writing down a numerical algorithm is to apply each contribution to a whole vector or matrix of values.
However, this is not the optimal way to implement the algorithm with respect to the number of intermediate values that are alive at a given time.
Each individual entry in the vector or matrix needs to be alive until all contributions are applied, and the entry can be used.
Instead, computing all contributions on a single entry and then immediately using and discarding that computed value requires only a single value to be alive at the same time.
This requires altering the order in which contributions are calculated.

%
%



Compilers already do unroll simple loop nests and reschedule their arithmetic expressions for minimal register usage, but refrain from doing so for larger, more complicated loop nests either because of heuristics or because of other impediments.

Another example of algorithmic restructuring is the computation of the elemental RHSs, which were obtained by multiplying the elemental matrices by the unknowns from the previous time step.
The computation of the elemental matrices is a hold over from a previous time when Alya still used implicit time-stepping, which required the assembly of a global matrix.
It is possible to obtain exactly the same elemental RHSs directly, without the need to assemble intermediate elemental matrices.
Instead of computing the large number of entries in the elemental matrix, values can individually be computed, used and discarded one by one, which reduces the number of intermediate results that are alive at the same time.

\subsection{Algorithmic Specialization} \label{Specialization}

The original assembly subroutine takes function parameters that specify the kind of element and the number of nodes and Gauss integration points.
We specialize the assembly subroutine for linear tetrahedral elements, which makes the number of four nodes per element and four Gauss integration points compile time parameters.
Moreover, the gradients of the shape functions are constant for tetrahedral elements.
This means that the gradients are the same at all Gauss points, contrary to what happens for other elements.
Additionally to saving some computations, this also reduces the amount of intermediate values.

When different numerical treatments are available, we select the most used ones by setting the corresponding flag to a Fortran parameter instead of leaving it as a run time variable read from the input files. This avoids if statements and helps the compiler to optimize better. 

In nearly all of the flow problems we solve, density and viscosity are constant. However, the default version of Alya also enables it to run with non-constant values that may depend on some other variables, such as the temperature.
Therefore, specific subroutines calculate the density and viscosity depending on the constitutive model that the user selects in input files.
We only allow for constant density and viscosity in the current implementation and set them as Fortran parameters. 

The user can choose from several different turbulence models in Alya's default implementation. Instead, we prefer to include only the Vreman turbulence model, which we use most.
In the default Alya, turbulent viscosity is obtained at the beginning of each time step in a specific subroutine.
Instead, it is much more efficient to calculate it directly on the fly when performing the assembly. Finally, since the Vreman turbulent viscosity depends on the velocity gradients, constant for tetrahedral elements, we do not need to obtain one value for each Gauss point but only one value per element.
This saves arithmetic operations as well as intermediate values.

The version \textbf{RS} includes both \textbf{R}estructuring for minimized temporary value lifetime from Section~\ref{lifetimeopts} as well as the \textbf{S}pecializations described in section \ref{Specialization}.
The combination of both measures reduces the number of temporary values from 430 values per element in 32 arrays to 130 values per element in 13 arrays.

\paragraph{GPU Results}
The number of floating point operations (see column \textbf{RS} in Table~\ref{tab:metrics}) decreases by $4\times$, both because some terms need not be computed any more in the specialized case, as well as due to the elimination of redundant computations.
These measures, as well as the structure changes aimed at minimizing the lifetime of intermediate values, also reduce the number of load and store operations by $6\times$ compared to the baseline version \textbf{B}.
The lower total data volume increases the effectiveness of both caches to $60\%$, which accumulates to a total reduction in DRAM data volume and matching speedup of $20\times$ compared to the baseline version \textbf{B}.
While it does not increase occupancy, the lower register accumulation of 184 is a sign of fewer intermediates.

\paragraph{CPU Results}
Just as on the GPU, version \textbf{RS} executes about $3\times$ fewer floating point operations (see column \textbf{RS} in Table~\ref{tab:metricscpu}) compared to the baseline version \textbf{B} on the CPU.
At the same time, the reduced intermediate value data volume increases the L1 cache effectiveness to $94\%$ and the L2 cache data volume is reduced by about $10\times$.

The DRAM volume decreases only by a small amount, because the L2 and L3 caches are already effective even in the unoptimized version to serve most of the redundant traffic.
The version \textbf{RS} is $3\times$ faster on a single core, which is a similar ratio as the reduction in floating point operations.

\subsection{Improvements: Privatization} \label{LocalMemory}

The explicitly allocated intermediate value arrays can be replaced by thread private scalars by dropping the \lstinline[style=f90]{VECTOR_DIM} dimension in the array declaration as well as in the addressing and by adding the \verb!private! clause in the OpenACC statement, as has been done in the test code in Listing~\ref{lst:private}.

On GPUs, using the OpenACC \verb!private! clause causes the compiler to map the value into \emph{local memory} instead of global memory.
We compile and analyze three different versions of the test code in Listing~\ref{lst:private} to analyze what this means for code generation and execution on the GPU.
The compiler maps \lstinline[style=f90]{temp} either to:

\begin{enumerate}
\item Global Memory, because \lstinline[style=f90]{temp} is a global, 2D array.
\item Local Memory, because \lstinline[style=f90]{temp} is a private, 1D array.
\item Registers, because \lstinline[style=f90]{temp} is a private, 1D array, with a compile time constant \lstinline[style=f90]{rowlen}. After unrolling the loops, the array indices are compile time constants too and the compiler can map the variables to registers.
\end{enumerate}

Table~\ref{tab:privatestore} shows the number of executed store instructions per thread as well as the stored data volumes for each version.
The assignments to \lstinline[style=f90]{temp} in the first loop nests are realized by the compiler with 8 global stores in the first case (global memory), 8 local stores in the second (local memory), and no store instructions at all in the third case (registers).

\begin{lstfloat}[t]
\begin{lstlisting}[style=f90]
integer, parameter :: rowlen = 8
real(kind=8) :: temp([-VECTOR_DIM,-] rowlen)

[/!$acc parallel loop/] [!private(temp)!]
do ivect = 1,VECTOR_DIM
   do row=1,rowlen
      temp([-ivect,-] row) = row * A(ivect)
   end do

   B(ivect) = 0
   do row = 1, rowlen
      B(ivect) = B(ivect) + temp([-ivect,-] row)
   end do
end do
\end{lstlisting}
  \caption{ Test code that shows the changes through Privatization of temporary arrays}\label{lst:private}
\end{lstfloat}
The measured stored data volumes in Table~\ref{tab:privatestore} show that both local and global are written through to the L2 cache, but only global store are always written back to DRAM.
Instead, cache lines modified by local stores can just be invalidated when the thread they belong to finishes execution, as long as they are not evicted before that due to insufficient capacity.
This does not happen in the test code with its small L2 cache footprint, but frequently with the full application.

In our optimizations of Alya, the specialization to a certain number of dimensions (2D or 3D), the element types and number of Gauss points makes the loop lengths of many small loops a compile time parameter, so that the compiler can then map the privatized intermediate results to registers, and only spill values to local memory if the number of registers is insufficient.

\paragraph{P: GPU Results}
Version \textbf{P} is a variant where the only change on top of the base line is the privatization of all arrays, which allows to study the isolated effects of the temporary array privatization.

Column \textbf{P} in Table~\ref{tab:metrics} shows the unsurprising conversion of many global loads and stores to local ones.
The total number of load and store operations halves, which indicates that the compiler was able to keep intermediates in registers more often and for longer.
The compiler still has to spill intermediates to local memory, because the register allocation is still maxed out.
The reduced load and store operation count is offset by decreased cache effectiveness, which is why the DRAM volume does not decrease much.
The lower cache effectiveness indicates that mostly redundant, well cacheable loads and stores were eliminated.
Because no algorithmic changes were made, the floating point operation count stays the same.
The roofline diagram in Figure~\ref{fig:roofline} shows that while the arithmetic intensity does not change much, the memory bandwidth doubles and is now fully utilized.
The reason for the more than $2\times$ speedup despite the similar floating point operation count, data volumes, and low occupancy, is fewer memory access stalls and generally decreased instruction counts.

\paragraph{RSP: GPU Results}

\renewcommand\theadalign{br}
\renewcommand\theadfont{\bfseries}
\renewcommand\theadgape{\Gape[4pt]}
\renewcommand\cellgape{\Gape[4pt]}

\renewcommand{\arraystretch}{1.8}

\begin{table}[t]
  \begin{tabular}{r r | c c c}
    \multicolumn{2}{r}{\thead{array temp  mapped to: }}&   \thead{global \\ memory} & \thead{local \\ memory} & \thead{registers} \\ \hline

    \rowcolor{myc0} \cellcolor{myw}  & local &      &  8 &  \\
    \rowcolor{myc5} \cellcolor{myw} \multirow{-2}{1.6cm}{\hfill\normalsize{store instructions}} &  global & 9   &  1  & 1   \\ \hline
    
  \rowcolor{myc0} \cellcolor{myw}  & to L2 &   $72\,B$   &  $72\,B$ & $8\,B$ \\
    \rowcolor{myc5} \cellcolor{myw} \multirow{-2}{1.2cm}{\hspace{-0.3cm}\normalsize{store data volume}} &  to DRAM & $72\,B$   &  $8\,B$  & $8\,B$    \\
    
  \end{tabular}
  \vspace{0.2cm}
  \caption{Type and number of store instructions and measured store data volumes per thread for the code in Listing~\ref{lst:private} on GPUs}\label{tab:privatestore}
\end{table}

The variant \textbf{RSP} builds on version \textbf{RS}, but converts the global vector intermediates to scalar, private arrays.

The measured data in column \textbf{RSP} in Table~\ref{tab:metrics} shows that compared to \textbf{RS}, this reduces the number of load and store operations by $8\times$, because the reduced amount of intermediates leads to less spilling than version \textbf{P} and fewer allocated registers.
The scattered, indirect memory accesses typical for sparse matrices are now the majority instead of the uniform, repeated accesses by the same thread due to spilling.
The L1 cache shows zero effectiveness for these loads.
In the device wide shared L2 cache, there is much more potential for reuse of nodal values loaded by different threads, and thus a higher effectiveness.

Due to fewer restrictions for the compiler and despite no algorithmic changes, the floating point instruction count decreases by about $20\%$.
Fewer load/store instructions also leads to fewer address calculation instructions, as well as to fewer memory stalls.
A smaller register allocation increases occupancy.
Combined, these factors lead to a speedup of more than $2\times$.

\paragraph{RSP: CPU Results}
The CPU path of \textbf{RSP}, resembles the structure of the GPU / OpenACC path: the Fortran array expressions are replaced with an enclosing loop that is annotated with a \lstinline[style=f90]{!$omp  simd private (temp)} instead of the OpenACC statements similar as in Listing~\ref{lst:private}.
The OpenMP clause is only used as a portable SIMD hint to the compiler, and not strictly necessary.
Alya is a pure MPI code and OpenMP is not used for core level parallelism.
We can only speculate whether the vectorized Fortran style using array slices had been necessary for vectorization in the past, and has been made obsolete in this case through advancements in the compilers' ability to vectorize loops.

\begin{figure}
  \centering
  \includegraphics[width=\columnwidth]{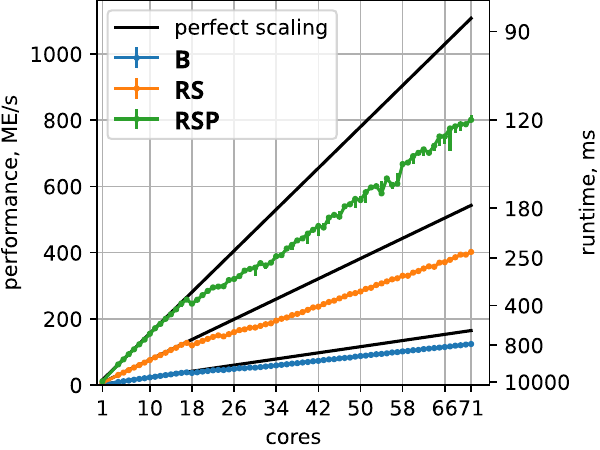}
   \caption{Performance in mega elements per second (left axis) and wall time in ms (right axis, inverted) vs CPU core count excluding Alya's master process. Perfect scaling line extrapolated from 4 worker rocesses. Errorbars indicate the lowest, highest and median value out of three runs.}
   \label{fig:cpuscaling}
 \end{figure}
The CPU path still has a separate, unvectorized loop for the scattering of the local matrix to the global RHS to avoid lost updates in the reduction of the global RHS entries.
While the CPU and GPU paths are implemented as two different sources, the two paths are more similar than before, and returning to a single unified source is possible.

The measured data in column \textbf{RSP} in Table~\ref{tab:metricscpu} shows a drop in the number of floating point operations by a third, to a similar level as the same variant on the GPU, again, because of fewer restrictions on the compiler in terms of intermediate expressions.
Together with a slight increase in the floating point operation throughput, this change results in a $80\%$ single core speed up.

Figure~\ref{fig:cpuscaling} shows linear scaling of all three CPU version up to 17 worker processes + 1 master process, beyond which the turbo modes selects lower frequency bins of \unit[3.1]{GHz} and later \unit[2.6]{GHz} instead of the maximum \unit[3.4]{GHz}, which comes with an exactly proportional slowdown visible in the graph (see Intel's Turbo Tables \cite{turbotables}).
At only \unit[2.5]{GB/s} DRAM transfer rate for a single core, even using all 36 cores the fastest variant \textbf{RSP} does not saturate the single socket memory bandwidth, so linear scaling is expected.

\subsection{Results: Further GPU Lifetime Optimizations}
On the GPU, additional restructuring can reduce register count to 128 registers without spilling.
The largest part is the immediate scattering of local RHS entries to the global matrix instead of first computing the entire local RHS.
This version, called \textbf{RSPR}, is not transferable to the CPU, as it breaks the concept of a single vectorization loop and a scalar scatter loop.

The code structure changes lead to fewer intermediate values, which is reflected in less spilling of data and fewer allocated registers (see column \textbf{RSPR} in Table~\ref{tab:metrics}).

Less spilling increases L2 cache effectiveness, which decreases the data volumes further.
Performance increases by about 30\%, which to a large part is explainable by the increase in occupancy by 33\% due to the lower register allocation.

\begin{figure}[t]
  \centering
    \includegraphics[width=\columnwidth]{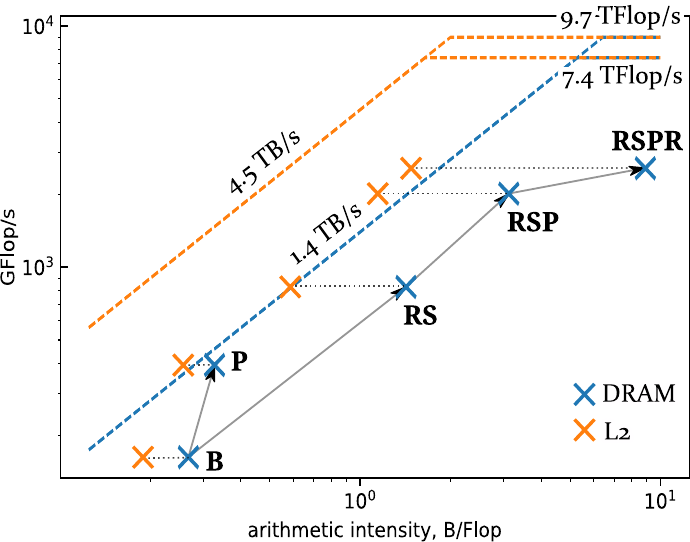}
    \caption{Roofline diagram of the GPU versions DRAM and L2/s. Measured bandwidths from \cite{cudabenches}. Includes a lower roof of \unit[7.4]{TFlop/s} due to the application instruction mix.}
    \label{fig:roofline}
\end{figure}

The roofline diagram illustration in Figure~\ref{fig:roofline} of the data points from Table~\ref{tab:metrics} shows how the arithmetic intensity continually increases with the consecutive variants, up to the point that the last version \textbf{RSPR} is past the roofline knee of the A100 GPU.
The L2 cache arithmetic intensity does not increase as much as the DRAM arithmetic intensity and more optimized versions are rather L2 cache bandwidth limited ($41\%$ of peak) than floating point execution ($32\%$ of peak) limited.
This is because the essential, remaining memory accesses consist of reading of nodal values, and reductions of RHS entries, which both are shared and reused by multiple threads in the L2 cache.






\section{Energy Efficiency}\label{sec:energy}

We estimate the power consumption of one \emph{Alex} GPU including its share of the host system power consumption and one \emph{Fritz} node from their TOP500 \cite{top500} entries by dividing the system power values by the GPU count or node count.
At \unit[51]{ms} (kernel time only) and \unit[421]{W}, the fastest GPU version consumes \unit[21]{J}.
By the same estimate, a full \emph{Fritz} node with a consumption of \unit[683]{W} for \unit[122]{ms} consumes a minimum of \unit[82]{J}.

Executing these particular implementations of the element assembly routine is about $4\times$ more energy efficient on one A100 GPU in \emph{Alex} compared to the CPU cores in a full \emph{Fritz} node.
This is in a similar region as an energy efficiency estimate based on the two system's Green500~\cite{green500} entries, which has the GPU system about $5\times$ more efficient.

For the baseline version, where execution on the GPU is $4-5\times$ slower than on the CPUs, using the GPU system would have been the less energy efficient option.




\section{Conclusions}

In this paper we have presented strategies on how to optimize the OpenACC port of the RHS assembly in the existing Fortran code Alya by more than $50\times$.
In recent years, the high productivity of annotation-based GPU programming approaches has made it possible to tackle the big challenge of converting large codes to the GPU.
We believe that the style of code found in Alya is not uncommon for such initial ports of CPU codes and that similar measures can be applied to other code bases.

In the course of the optimizations we initially challenged the feasibility of a performance-portable unified code base and created a dedicated source version for the GPU.
We believe it is worthwhile to work with separate versions during the development process to obtain the optimal implementation, and be able to quantify the performance loss of portability.
We found that the penultimate GPU version could be easily unified with the last CPU optimization, which would result in a unified version that would be simpler and more productive than the baseline unified implementation.

The energy efficiency on GPUs compared to CPUs obtained by Alya is now very close to what one could expect from the energy efficiency ratios of CPU and GPU supercomputers from the Top500 list.
Even though it is easy to find recent literature on high-performance GPU implementations of high-order methods, this is not the case for linear finite elements used in this work.

Part of the improvements has been possible thanks to specialization. Thus, our current implementation can not cover the full range of problems the original code could handle. However, it can deal with a large variety of problems of industrial interest.
Alya's programming paradigm has usually favored a unified implementation for a broader range of problems rather than developing specialized versions.
This work challenges such an approach by showing the benefits of specializing to a smaller range of problems.

So far we concentrated on optimizing momentum right-hand-side assembly, which is the most costly kernel of large-eddy simulations solved with an explicit time discretization.
This step is trivially parallel; we have therefore not devoted efforts to testing the scalability.
For the kind of incompressible flow problems we are interested in, the solution of the linear system for the pressure is the second most computationally expensive kernel and the main challenge for scalability.
Future work will focus on finding solvers with the correct algorithmic scalability for exascale hardware and porting of a few remaining parts of the code that currently still cause host-device memory transfers.

Alya's sources are available for non-commercial purposes. Contact the first author.


\bibliographystyle{IEEEtran}
\bibliography{references}

\end{document}